\documentclass[amssymb,useAMS,prd,aps,amsmath,superscriptaddress,,nofootinbib]{revtex4}
\usepackage{pslatex}
\usepackage{graphicx}
\usepackage{psfrag}

\begin{document}
\newcommand {\be}{\begin{equation}}
\newcommand {\ee}{\end{equation}}
\newcommand {\ba}{\begin{eqnarray}}
\newcommand {\ea}{\end{eqnarray}}
\title{ A FRAMEWORK TO SIMULTANEOUSLY EXPLAIN\\
TINY NEUTRINO MASS AND HUGE MISSING MASS PROBLEM OF THE UNIVERSE }
\author{YASAMAN FARZAN}

\address{School of Physics, Institute for Research in Fundamental Sciences
(IPM) \\
Tehran, PO. Box 19395-5531, Iran\\
yasaman@theory.ipm.ac.ir}



\begin{abstract}

Recently a minimalistic  scenario has been developed to explain
dark matter and tiny but nonzero neutrino masses. In this
scenario, a new scalar called SLIM plays the role of the dark
matter. Neutrinos achieve Majorana mass through a one-loop
diagram. This scenario can be realized for both real and complex
SLIM. Simultaneously explaining the neutrino mass and dark matter
abundance constrains the scenario. In particular for real SLIM, an
upper bound of a few MeV on the masses of the new particles  and a
lower bound on their coupling are obtained which make the scenario
testable. The low energy scenario can be embedded within various
$SU(2)\times U(1)$ symmetric models. I shall briefly review the
scenario and a specific model  that embeds  the scenario, with
special emphasis on the effects in the charged Kaon decay which
might be observable at the KLOE and NA62 experiments.

\end{abstract}
\maketitle

\section{Introduction}
The standard models of particles and cosmology have been
unexpectedly successful in explaining various observed phenomena;
however, there are   two perplexing puzzles that are still
unsolved: tiny but nonzero neutrino masses and the nature of dark
matter. A considerable number of models beyond the Standard Model
(SM) have been developed that address each of these puzzles but
attempts to simultaneously explain them have only recently been
started. In this letter, we shall review an economic scenario
proposed in Ref.~\cite{scenario} which simultaneously explains the
neutrino mass and dark matter puzzles. In this scenario, neutrinos
acquire mass at one-loop level and the Dark Matter (DM) particles
are produced thermally.  The information on the neutrino mass
matrix combined with the information on the dark matter abundance
constrain the scenario and in most cases, make it testable.

The scenario is based on an effective low energy Lagrangian which
has to be embedded within an electroweak symmetric model. The
features of scenario show a guiding path to build models that
embed the scenario. In  Refs.~\cite{myModel,ourModel}, two
distinct models have been developed that embed the scenario. We
review the model introduced in Ref.~\cite{myModel} and discuss its
predictions. The model inherits the testability from the scenario.
Moreover, the new fields added to make the model $SU(2)\times
U(1)$ invariant cause observable effects in the LHC as well as in
the low energy experiments searching for lepton flavor violation
such as $\mu \to e \gamma$ searches.  Since the model is very
economic, as we shall see the information from these various low
energy and high energy experiments can provide a possibility for
cross check of the model.

The letter is organized as follows. In section~\ref{scenario}, we
 introduce the content of the scenario and discuss the neutrino
mass production and DM annihilation. In section \ref{phenom}, we
review the observable effects in the meson decay, core collapse
supernova explosion and indirect DM searches with special emphasis
on the testability of the model. In section \ref{model}, we review
the model introduced in Ref.~\cite{myModel} and review the
predictions of the model for the LHC and lepton flavor violating
rare decays. In section \ref{discussion}, we summarize the results
and discuss the prospects.

\section{The scenario \label{scenario}}

The scenario requires adding only the following particles to the
SM:  (i) a scalar which we call it SLIM: $\phi$; (ii) Two or more
right-handed Majorana neutrinos, $N_i$ with mass term $${m_{N_i}
\over 2} N_i^T (i\sigma_2) N_i \ . $$
 These new particles couple
to the SM neutrinos through \be \label{mother} \mathcal{L}=g_{i
\alpha} \phi \bar{N}_i \nu_\alpha .\ee Of course, such a coupling
not being invariant under the $SU(2)\times U(1)$ symmetry is only
a low energy effective coupling and at higher energies, as we
shall discuss in section \ref{model} has to be augmented to become
electroweak symmetric. A $Z_2$ symmetry under which only new
particles are odd is imposed which forbids a Dirac mass term of
type $\bar{N}_i \nu_\alpha$ and makes the lightest new particle
stable and a suitable candidate for DM. We take $\phi$ to be the
lightest new particle and therefore the DM candidate. The scenario
can be realized both for cases that the scalar is real and
complex. Let us study them one by one.
\subsection{Real SLIM}
\begin{figure}[h]
  \includegraphics[width=5.5cm,bb=60 120 338
290,clip=true]{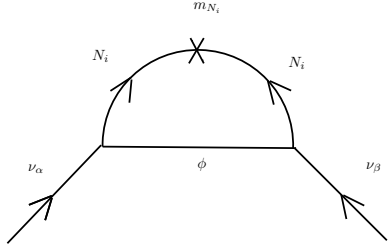}
  \caption{ Neutrino mass
diagram} \label{nuMass}
\end{figure}

Real SLIM cannot carry lepton number. Neither do $N_i$ which are
of Majorana type. Thus, the coupling in Eq.~(\ref{mother})
violates lepton number conservation. Within this scenario,
neutrinos through the one-loop diagram shown in Fig.~\ref{nuMass}
acquire Majorana masses that can be written as follows
\begin{equation}
\label{flavoredmass} (m_\nu)_{\alpha \beta} =\sum_i \frac{g_{i
\alpha}g_{i \beta}}{16
  \pi^2}m_{N_i}\left(\log
  \frac{\Lambda^2}{m_{N_i}^2}-\frac{m_\phi^2}{m_{N_i}^2-m_\phi^2}\log
\frac{m_{N_i}^2}{m_{\phi}^2} \right),
\end{equation}
where $\Lambda$ is the ultraviolet cutoff of the effective
coupling which is the electroweak scale. The neutrino mass matrix
can in general be parameterized as $$m_{\nu}=U \cdot {\rm
Diag}[m_1,m_2 e^{2i \gamma_2},m_3e^{2i\gamma_3}] U^T,$$ where $U$
is the famous PMNS mixing matrix and $\gamma_2$ and $\gamma_3$ are
Majorana CP-violating phases. To accommodate the neutrino flavor
structure, the coupling has to have the following structure:
\begin{equation}
  \label{PhiRealComplex} g={\rm Diag}(X_1,..., X_n)\cdot
  O \cdot {\rm Diag}(\sqrt{m_1},\sqrt{m_2}e^{i \gamma_2},\sqrt{m_3}e^{i
    \gamma_3}) U^T \ ,
\end{equation}
where  $n$ is the number of Majorana neutrinos  and
\begin{equation}
\label{real-Xi} X_i=4\pi
\left(\frac{1}{m_{N_i}}\right)^{1/2}\left( \log
\frac{\Lambda^2}{m_{N_i}^2} -{m_\phi^2 \over m_{N_i}^2-m_\phi^2}
\log \frac{m_{N_i}^2}{m_\phi^2} \right)^{-1/2},
\end{equation}
where $O$ is an arbitrary $n \times 3$ matrix that satisfies $O^T
\cdot O={\rm Diag}(1,1,1)$. Notice that for $n=3$, this means $O$
is an arbitrary orthogonal matrix. To accommodate the neutrino
data with two nonzero mass eigenvalues, at least two nonzero
right-handed neutrinos are needed. With only two right-handed
neutrinos, one of the mass eigenvalues will vanish and the
neutrino mass scheme will be hierarchical.

In this scenario, the DM in the early universe is thermally
produced so its abundance is given by the inverse of its
annihilation cross-section; see Ref.~\cite{leeweinberg}.  The main
annihilation modes are lepton  number violating ones $\phi\phi \to
\nu \nu, \bar{\nu} \bar{\nu}$ which proceed via a $t$-channel
lepton number violating right-handed neutrino exchange ($\langle N
N\rangle$):
\begin{equation} \label{sigma}
\langle \sigma(\phi \phi \to \nu_\alpha \nu_\beta ) v
\rangle=\langle \sigma(\phi \phi \to \bar{\nu}_\alpha
\bar{\nu}_\beta ) v \rangle=\frac{1}{4\pi}\left|\sum_i {g_{i
\alpha } g_{i\beta } m_{N_i}\over m_\phi^2+m_{N_i}^2}\right|^2.
\end{equation}
To account for the observed DM abundance , the annihilation cross
section has to be of order of $10^{-36}~{\rm cm}^2$. Setting
$\langle \sigma_{tot} v\rangle$ equal to this value and inserting
the observed neutrino masses, we find that the lighter
right-handed neutrinos dominating the DM annihilation should have
a mass in the range $$ O(1)~{\rm MeV}\lesssim m_{N_1} \lesssim
10~{\rm MeV}.$$ The rest of $N_i$ can be heavier. To prevent the
decay of the SLIM, it has to be lighter than $N_1$.  From this
consideration, we do not obtain any lower bound on the SLIM mass.
The large scale structure arguments impose a lower bound of a few
keV on the SLIM mass; see Refs.~\cite{WDM1,WDM2,WDM3,WDM4}.
According to Ref.~\cite{serpicoraffelt}, Nucleosynthesis imposes
even a stronger lower bound of MeV  so \be O(1)~{\rm MeV}\lesssim
m_\phi< M_{N_1}\lesssim 10 ~{\rm MeV}.\ee Thus, SLIM has to be in
the MeV range. In fact, SLIM stands for Scalar as LIght as MeV.
From Eq.~(\ref{sigma}), we find
\begin{equation}
\label{gvalues} g_{1 \alpha} \simeq 10^{-3} \
\sqrt{\frac{m_{N_1}}{10 \ \rm{MeV}}} \ \left(\frac{\langle \sigma
v_r\rangle}{10^{-26} \rm{cm^3/s}}\right)^{\frac{1}{4}}\left({1+
\frac{m_\phi^2}{m_{N_1}^2}}\right)^{\frac{1}{2}}
\end{equation}
which leads to
\begin{equation}
\label{greal} 3 \times 10^{-4} \lesssim g_{1 \alpha} \lesssim
10^{-3}.
\end{equation}
Lightness of $N_1$ and $\phi$ and the lower bound on their
coupling imply that the scenario is testable in low energy
phenomena. We will discuss this in more detail in  section
\ref{phenom}.
\subsection{Complex SLIM \label{complex}}
A complex $\phi$ can be decomposed in terms of its real components
as $\phi \equiv (\phi_1+i\phi_2)/\sqrt{2}$.  In practice, we have
now two scalars with couplings $g_{i \alpha}/\sqrt{2}$ and $ig_{i
\alpha}/\sqrt{2}$. The mass term for $\phi$ can in general be
written as \be \label{Vm} \mathcal{V}_m= M^2 \phi^\dagger \phi
-\frac{m^2\phi \phi +{\rm H.c.}}{2}.\ee Imposing CP symmetry on
$\mathcal{V}_m$ implies that $m^2$ is real. In this case, $\phi_1$
and $\phi_2$ will be mass eigenstates with masses
\begin{eqnarray}
m_{\phi_1}^2 &=& M^2-m^2 \\
m_{\phi_2}^2 &=& M^2+m^2.
\end{eqnarray}
Thus, for nonzero $m^2$, there is a splitting between the masses.
Neutrinos obtain Majorana masses through the one-loop diagram
shown in Fig.~\ref{nuMass}. Diagrams with $\phi_1$ and $\phi_2$
propagating in the loops should be summed up. Since their
couplings are the same up to a factor of $i$, the cutoff dependent
parts of the contributions of  $\phi_1$ and $\phi_2$ cancel each
other and we arrive at the following formula
\begin{eqnarray}
\label{flavorcomplexmass}
 (m_{\nu})_{\alpha \beta} =  \sum_i \frac{g_{i\alpha}g_{i\beta
 }}
 {32 \, \pi^2} \ m_{N_i}  \left[\frac{m_{\phi_1}^2}{(m_{N_i}^2-
 m_{\phi_1 }^2)}
\ln\frac{m_{N_i}^2}{m_{\phi_1}^2}-
\frac{m_{\phi_2}^2}{(m_{N_i}^2-m_{\phi_2}^2)}
\ln\frac{m_{N_i}^2}{m_{\phi_2}^2} \right].
\end{eqnarray}
Notice that for $m_{\phi_1} \to m_{\phi_2}$, $m_\nu$ vanishes.
This is understandable because at this limit ({\it i.e.,} $m^2 \to
0$), the Lagrangian becomes lepton number conserving by assigning
a lepton number of $-1$ to $\phi$. As a result, in this limit, the
lepton number violating Majorana mass, $m_\nu$, should vanish.

Without loss of generality, we can take $\phi_1$ to be lighter
than $\phi_2$ and therefore a DM candidate. Since the couplings of
$\phi_1$ are similar to real SLIM up to a factor of $\sqrt{2}$,
the annihilation cross section of $\phi_1$ is given by
Eq.~(\ref{sigma}) replacing $g$ with $g/\sqrt{2}$.

Unlike the real SLIM case, in the complex scenario,  the
right-handed neutrinos can be as heavy as the electroweak scale or
even heavier. $\phi_1$ and $\phi_2$ can be heavy, too. For
$m_{\phi_1}\gg 10$~MeV, $\phi_1$ and $\phi_2$ will be
quasi-degenerate and their coannihilation at freeze out has to be
taken into account. Notice however that within this scenario,
$\sigma (\phi_1 \phi_2 \to \nu \nu)\sim \sigma (\phi_1 \phi_1 \to
\nu \nu)$ so the overall picture will not be altered by
coannihilation.

\section{Phenomenological implications of the scenario\label{phenom}}

The scenario, specially in the real SLIM mode, contains light
particles with small (but bounded from below) coupling with $\nu$.
As a result, we expect the scenario to  have its imprint in
various phenomena. One possibility is nucleosynthesis
 which as shown in Ref.~\cite{serpicoraffelt} implies a lower bound of MeV on the
SLIM mass.
 Different possibilities have been enumerated in
 Ref.~\cite{scenario}. In this section, we discuss the effects in particle
 decays, indirect searches by neutrino detectors as well as core collapse
 supernova explosion which are
 among the most promising observations to look for the SLIM.
\subsection{Light Meson and tau lepton decays}
Consider any of the particle decays within SM that produce a
neutrino which appears as missing energy: \be \label{AtoBNu} A \to
B +\nu.\ee In the presence of the new coupling Eq.~(\ref{mother}),
along with this decay mode, $A$ can go through the following decay
via a virtual $\nu$ exchange \be \label{AtoBNPhi}A \to B\ N_i \phi
\ee of course provided that the difference between $A$ mass and
the sum of masses in $B$ is larger than the sum of the masses of
$\phi$ and $N_i$. Remember that in case of real SLIM, we find
$m_\phi < m_{N_1}<10~{\rm MeV}$ so decays such as $K(\pi)\to
\mu(e) N_1 \phi$ or $\tau \to \mu (e) \nu N_1 \phi$ are possible.
$\phi$ and $N_i$ will show up as missing energy. Thus, by
comparing the measured value of $A \to B+{\rm missing~energy}$
with the SM prediction $|g_{i \alpha}|$ can be constrained. Such
an analysis has not been performed for the SLIM model but similar
analysis has extensively been carried out (see
Refs.~\cite{Britton:1993cj,Barger:1981vd,Gelmini:1982rr,Lessa}) in
the case of the Majoron couplings to neutrinos $\tilde{g}_{\alpha
\beta} J \bar{\nu_\alpha^c} \nu_\beta$ where $J$ is the Majoron
which is a massless pseudoscalar. Such a coupling will also
contribute to $A\to B+{\rm missing~energy}$. As long as the sum of
the masses of $N_i$ and $\phi$ is much smaller than the difference
between the masses of $A$ and $B$, the bounds for Majoron also
apply for the SLIM case. The strongest bounds come from the Kaon
and tau decay for which this condition is satisfied.

The new SLIM interaction can give rise to new channels of ($K^+
\to \mu^++{\rm missing~ energy}$) and ($K^+ \to e^++{\rm missing~
energy}$) causing a deviation from the universality predicted by
the SM: \be \label{universality} {\Gamma(K^+\to e^+ +{\rm missing~
energy}) \over\Gamma(K^+\to \mu^+ +{\rm missing~ energy})}={
\Gamma_{SM}(K^+ \to e^+ \nu_e) +\sum_i\Gamma(K^+ \to e^+ N_i \phi)
\over \Gamma_{SM}(K^+ \to \mu^+ \nu_e) +\sum_i\Gamma(K^+ \to \mu^+
N_i \phi)}\ee
$$ \simeq {
\Gamma_{SM}(K^+ \to e^+ \nu_e) +\sum_i\Gamma(K^+ \to e^+ N_i \phi)
\over \Gamma_{SM}(K^+ \to \mu^+ \nu_e)}.$$ By comparing the SM
prediction with the most recent results from KLOE in
Ref.~\cite{Kloe}, one can obtain
$$ \sum_i |g_{ie}|^2<10^{-5}$$ where the sum runs over $N_i$
lighter than $K^+$. Notice that this bound is too weak to probe
the range in Eq.~(\ref{greal}). Another bound can be obtained by
studying the spectrum of the muon produced in the Kaon decay. Of
course, the spectrum of the muon in $K^+ \to \mu^+ N_i \phi$,
being a three-body decay, will be considerably different from that
in $K^+ \to \mu^+ +\nu_\mu$, being a two-body decay. An indirect
way to extract the bound is to use the bound on $K^+ \to \mu^+
\nu_\mu \nu \bar{\nu}$. This four-body decay is allowed in the SM
but the SM prediction is completely negligible and far below the
present bound  found by Ref.~\cite{73} which is \be \label{4body}
{\rm Br}(K^+ \to \mu^+ + \nu_\mu+\nu+\bar{\nu})<6 \times 10^{-6} \
.\ee In Ref.~\cite{Lessa}, this bound is interpreted as the bound
on the new modes of $(K^+\to \mu^+ +{\rm missing~ energy})$. Such
an interpretation  is not of course completely accurate as the
spectrum for three body decay and four body decays are not the
same but tentatively it is acceptable. Taking the same approach,
we find \be \sum_i |g_{i \mu}|^2<9 \times 10^{-5}\ . \ee Notice
that this bound is based on the old LBL Bevatron data obtained in
1973. {It seems if the recent KLOE data is analyzed to search for
the $K^+ \to \mu^+ N_i \phi$ signal, a considerable part of the
range in Eq.~(\ref{greal}) can be probed.} Moreover upcoming NA62
experiment at CERN might be sensitive to even smaller values of
couplings. For a description of NA62 see Ref.~\cite{NA62}. Another
bound which has been studied in Ref.~\cite{Lessa} is the bound
from the $\tau$ decay which in our scenario can be translated into
$$\sum_i |g_{i\tau}|^2<10^{-1} \ . $$

In summary, the present bounds from meson and tau decays are too
week to probe the range in Eq.~(\ref{greal}); however,  there is a
hope that signals for this scenario already lie in the new data
from KLOE.
\subsection{Supernova type II explosion}

In our scenario, SLIMs can be produced by neutrinos inside the
supernova core. The produced SLIMs will undergo scattering on the
neutrinos already present in the core.  The scattering cross
section in the outer core where the  chemical potential of
neutrinos vanishes is given by
$$ \sigma(\phi \nu \to \phi \nu)\sim {g^4 T^2 \over 4 \pi
(T^2+m_{N_i}^2)^2}.$$ Taking the temperature to be of $T\sim 30$
MeV, we find that the mean free path is $(\sigma n_\nu)^{-1}\sim
10$~cm which is far smaller than the supernova core. This means
SLIMs will be trapped inside the core and their contribution to
cooling is only as a new single scalar mode diffuse out which can
be well tolerated within the present observational and the
supernova model uncertainties. Notice that in our scenario SLIMs
do not interact with nuclei so the constraint in Ref.~\cite{fayet}
does not apply to this scenario. For a more detailed discussion
see Ref.~\cite{scenario}.

It is also interesting to note that in the case of future
supernova neutrino observations, one may be able to test this
scenario by studying the neutrino energy spectra.
\subsection{Indirect DM searches}
In this scenario DM particles only interact with neutrinos so we
do not expect any photon signal from their annihilation (neither
prompt nor secondary). Moreover unlike WIMPs, SLIMs do not
interact with nuclei in the Earth or Sun so they do not get
trapped in them and we do not therefore expect a detectable
neutrino flux from DM annihilation in the Sun or Earth center.
However, the DM annihilation in the halo can result in a sizeable
neutrino flux at neutrino detectors such as super-kamiokande. This
effect has been studied in Ref.~\cite{silvia}. The present bound
from super-kamiokande is too weak to be sensitive to $\langle
\sigma_{tot} v \rangle \sim 10^{-36}$ cm$^2$. However, as shown in
Ref.~\cite{silvia}, with LENA (described in Ref.~\cite{LENA}) or a
megaton water detector doped with Gd, it will be possible to probe
the scenario for $m_\phi\sim 20-30$~MeV. Remember that such a mass
range is above the upper bound for real SLIM and can be realized
only for a complex SLIM.

\section{Models embedding the scenario\label{model}}

In Ref.~\cite{scenario}, several routes have been suggested to
embed the scenario in an electroweak invariant model. As we saw in
section~ \ref{scenario}, in the case of real SLIM, at least one of
the right-handed neutrinos has to be lighter than 10~MeV. As a
result, it should be an electroweak singlet to avoid the bounds
from the invisible decay width of the $Z$ boson. However, in the
case of complex SLIM, $N_i$ can be heavier than $m_Z/2$ and can
therefore be a component of an electroweak multiplet.

The model introduced in Ref.~\cite{ourModel} can be considered a
realization of a model embedding the complex SLIM scenario. On the
other hand, the model in Ref.~\cite{myModel} is an example for
embedding real SLIM scenario. In the following, we describe the
minimalistic model introduced in Ref.~\cite{myModel} that has
interesting predictions for the LHC and the lepton flavor
violating rare decays.

To make the coupling in Eq.~(\ref{mother}) $SU(2)\times U(1)$
invariant, $\phi$ is promoted to an electroweak doublet
$\Phi=(\phi^0 ~~\phi^-)$. Of course, in this case $\phi^0$ which
carries nonzero hypercharge  has to be complex and can be
decomposed as $(\phi_1+i\phi_2)/\sqrt{2}$ in terms of real
scalars, $\phi_1$ and $\phi_2$. With a coupling of form ${\rm
Re}[(H^T(i\sigma_2) \Phi)^2]$, the vacuum expectation value of the
Higgs leads to a small splitting between the $\phi_1$ and $\phi_2$
components which is what desired in the complex SLIM model. At
first sight, it seems that with this minimalistic content, one can
embed complex SLIM scenario but this is not the case: $\phi_1$ and
$\phi_2$, being the components of an electroweak doublet have to
be heavier than about 80~GeV to satisfy the present bounds in
Refs.~\cite{LEP1,LEP2,LEP3}. On the other hand as we saw in
section \ref{complex}, within this scenario the splitting between
$m_{\phi_1}^2$ and $m_{\phi_2}^2$ has to be smaller than $(20~{\rm
MeV})^2$. That is $\phi_1$ and $\phi_2$ will be quasi-degenerate
and having coupling to the $Z$ boson, they can coannihilate in the
early universe leading to a too low dark matter density (see
Ref.~\cite{inert}). In Ref.~\cite{myModel}, a new scalar $\eta$,
singlet under the electroweak transformation, is added which mixes
with the neutral component of $\Phi$. The DM candidate, SLIM, is a
mixture of $\eta$ and $\phi_1$ with mixing angle $\alpha$.
Neutrinos obtain mass through one loop diagram. In fact, the
masses of the components of $\Phi$ play the role of cut-off in
Eq.~(\ref{flavoredmass}). The model inherits the features of the
real SLIM scenario ({\it i.e.,} lower bound on the coupling and an
upper bound on the masses of the SLIM and $N_1$) so the model is
testable by low energy high precision experiments such as the Kaon
decay measurements. Moreover, the charged component of $\Phi$
leads to observable effects on $(g-2)_\mu$ and lepton flavor
violating rare decays. The effects on $(g-2)_\mu$ is two orders of
magnitudes below the present sensitivity but as shown in
Ref.~\cite{myModel}, the present bound on $\mu \to e \gamma$
already constrains a part of the parameter space. The MEG
experiment searching for $\mu \to e \gamma$ can further constrain
the model or hopefully find a hint. If the masses of the
components of $\Phi$ are not too high, they can be copiously
produced via the electroweak interaction at the LHC. Their
subsequent decays can be driven by coupling in Eq.~(\ref{mother});
in particular, $\Gamma(\phi^-\to \ell_\alpha^- N_i) \propto |g_{i
\alpha}|^2$. That is by studying the decay modes of $\phi^-$, it
is possible to derive the flavor structure of $|g_{i \alpha}|$.
Remember that $g_{i \alpha}$ is the same coupling that determines
the neutrino mass matrix as well as the rates of lepton flavor
violating decays such as $\mu \to e \gamma$, $\tau \to  \mu
\gamma$ and $\tau \to e \gamma$. That is, information on new
couplings $g_{i \alpha}$ can be derived  by all these three
methods which provide a way for cross check and test of the model.

In Ref.~\cite{majid}, the discovery potential of the LHC is
explored taking into account the detailed background. The
production cross section is of order of a few hundred $fb$ for
$m_{\phi^-}<150$~GeV. It is shown in Ref.~\cite{majid} that by
employing the state-of-the-art cuts, signal significance can well
exceed $5~\sigma$ for the 14~TeV run of the LHC with 30 $fb^{-1}$.
The possibility of deriving the flavor structure of the coupling
$g_{i \alpha}$ is also discussed.

\section{Summary and discussion \label{discussion}}
In this letter, we have reviewed a scenario that links DM and
neutrino masses. The scenario is based on a scalar, $\phi$,
coupled to the SM left-handed neutrinos and two or more
right-handed Majorana neutrinos: $g_{i \alpha} \bar{N}_i
\nu_\alpha \phi$. The scalar plays the role of the DM and is
stabilized via a $Z_2$ symmetry which also forbids a Dirac mass
for neutrinos. DM particle in this scenario mainly annihilate into
$\nu \nu$ and $\bar{\nu}\bar{\nu}$. This would lead to a diffuse
flux from the DM halo. The present upper bound from
super-kamiokande is too weak to probe the scenario which is based
on the assumption of thermal production of the DM in the early
universe and therefore $\langle \sigma_{tot} v \rangle \simeq
10^{-36}~ {\rm cm}^2$. In future, a LENA type detector or a
megaton water detector doped with Gd can probe the model for DM
mass around 20 to 30 MeV. The scenario can be realized in two
cases: (i) real $\phi$; (ii) complex $\phi$.

In the real SLIM case, we found $m_\phi<m_{N_1}<10$ MeV and $3
\times 10^{-4}\lesssim g_{1 \alpha} \lesssim 10^{-3}$. As a
result, new missing energy modes are predicted at kaon and tau
lepton decays. The present bounds are too weak to probe the
scenario but eventual improvement can test the model because the
upper bound on $m_\phi$ and $m_{N_1}$ combined with the lower
bound on $g_{1 \alpha}$ imply a lower bound on the rates of the
new decay modes. A re-analysis of KLOE data with the aim of
searching for $K^+ \to \mu^+ \phi N_1$ and $K^+ \to e^+ \phi N_1$
will be most promising. Upcoming NA62 experiment might be even
more sensitive. Real SLIM, being light, can be also produced via
neutrinos in a supernova core. The produced SLIM will be trapped
in the core but will eventually diffuse out and contribute to the
cooling of the supernova core. This new cooling mode can be
tolerated within the present uncertainties but in future if the
observational and supernova model uncertainties are reduced, the
presence of this new mode can be tested.  The SLIM emission can
also distort the spectrum of the emitted neutrinos.

Complex SLIM can be as heavy as the electroweak scale but the mass
splitting between the CP-odd and -even components ({\it i.e.,}
$|m_{\phi_1}^2-m_{\phi_2}^2|$) must be between $(10~{\rm MeV})^2$
to $(20~{\rm MeV})^2$.

We discussed the possibility of embedding the scenario within a
$SU(2)\times U(1)$ invariant model. We focused on the model
developed in Ref.~\cite{myModel} which embeds the real SLIM
scenario. In addition to being falsifiable by low energy
experiments, the model predicts observable effects for lepton
flavor violating rare decays, $(g-2)_\mu$ and the LHC. In
particular, a hint for new physics in running searches at MEG is
expected.  The model has certain predictions for the LHC ({\it
see} Ref.~\cite{majid}). It is in principle possible to measure
$g_{i \alpha}$ at the LHC. This is the same coupling determining
the neutrino mass matrix and lepton flavor violating rare decays
and the flavor pattern of new modes of $K^+ \to \ell_\alpha^++{\rm
missing~energy}$. Thus, there is a possibility of cross-checking
the model by low energy and high energy experiments.

It might be possible to augment the scenario to account for the
baryon asymmetry of the universe through the mechanism introduced
in Ref.~\cite{lepto}.

\section*{Acknowledgments}
I would like to thank Celine Boehm, Thomas Hambye, Sergio
Palomares-Ruiz and Silvia Pascoli. I would also like to specially
thank Prof G. Isidori for useful remarks. I am also grateful to
the organizers of Planck 2009 meeting, where this talk was
presented, for their hospitality.

\end{document}